\documentstyle[12pt]{article}
\def\gsim{\stackrel{>}{{}_\sim}}
\def\lsim{\stackrel{<}{{}_\sim}}
\def\beq{\begin{equation}}
\def\eeq{\end{equation}}
\def\ul{\underline}

\def\half{\frac{1}{2}}
\def\ie{i.\ e.\ }
\def\etal{et al.\ }
\begin{document}
\begin{flushright}
VAND--TH--94--14--UPD\\
SHEP--95--08\\
hep--ph--9512229\\
December 1995
\end{flushright}
\begin{center}
\Large
{\bf Updated: Higgs Mass Bounds Separate Models\\
of Electroweak Symmetry Breaking}
\end{center}
\normalsize
\vskip.5cm
\begin{center}
{\Large Marco A. D\'\i az}\\
{\sl Physics Department\\
University of Southampton\\
Southampton, SO17 1BJ, U. K.}\\
\vspace{.5cm}
{\Large Tonnis A. ter Veldhuis} and
{\Large Thomas J. Weiler}\\
{\sl Department of Physics \& Astronomy\\
Vanderbilt University\\
Nashville, TN 37235, USA}\\
\vspace{.5cm}
\end{center}
\centerline{ABSTRACT}
Vacuum stability and metastability imply lower limits on the
mass of the higgs boson in the Standard Model (SM).
In contrast, we present an improved calculation of an upper limit
on the lightest higgs mass in supersymmetric (susy) models,
by  summing to all orders in perturbation theory the
leading and next--to--leading logarithms with a
renormalization group equation technique,
and by including finite two-loop QCD corrections.
The improvement lowers the Minimal Susy Standard Model (MSSM)
upper limit by about 10 GeV.
The main uncertainty in each limit is the value of the top mass,
which is now constrained by the recent Fermilab results.
We study the possibility
that these bounds do not overlap, and find that\\
(i) a mass gap emerges between the SM and the MSSM
at $m_t\sim 175$ GeV for $\alpha_s(M_Z^2)=0.118$
and at $m_t \sim 180$ GeV for more generous values $\sim (0.130)$ of
$\alpha_s$;
and between the SM and the Minimal plus Singlet Susy Model [(M+1)SSM] if
the independent scalar self--coupling of the latter is perturbatively small
or if the $\tan\beta$ parameter is large;
these gaps widen with increasing $m_t$; \\
(ii) the mass gap emerges with $m_t$ 10 GeV lighter if only vacuum stability
and not metastability is imposed;\\
(iii) there is no overlap between the SM and
the MSSM bounds at even smaller values of $m_t$ for
the $\tan\beta$ value ($\sim 1$--2) preferred in
Supersymmetric Grand Unified Theories.\\
Thus, a measurement of the first higgs mass will serve to exclude
either the MSSM/(M+1)SSM higgs sectors or the SM higgs sector.
In addition, we discuss the upper bound on the lightest higgs mass
in SUSY models with an extended higgs sector.
Finally, we comment on the discovery potential for
the lightest higgses in these models.
\vfill\eject
\section{Introduction}

The simplest and most popular
possibilities for the electroweak (EW) symmetry breaking
sector are the single higgs doublet of the minimal Standard Model (SM),
and the two higgs doublet sector of the Minimal Supersymmetric
Standard Model (MSSM).  Experimentally, very little is known about the
higgs sector of the electroweak model.  However, theoretically, quite a
lot of higgs physics has been calculated.
The electroweak symmetry--breaking scale is known:
the vacuum expectation value (vev) of the complex higgs field $\Phi$
is \linebreak $<0|\Phi|0>=v_{SM}/\sqrt{2}=175$ GeV.
This value is remarkably close to
the top quark mass of $176\pm 8\pm 10$ GeV
(which itself is very consistent with the values
inferred from precision electroweak data, assuming the SM:
$m_t=164\pm25$ GeV in \cite{ewgroup},
and more recently, $m_t=156\pm15$ GeV in \cite{ellisetal})
announced by the CDF collaboration at Fermilab \cite{CDF}.
Higgs mass bounds have been calculated, including loop corrections.
One aspect of the mass bounds \cite{kp} which we quantify
in this paper is the following:
inputing the CDF value for the top mass into quantum loop
corrections for the symmetry--breaking higgs sector leads to mutually
exclusive, reliable bounds on the SM higgs mass and
on the lightest MSSM higgs mass \cite{ourPRL,Casasetal}.  From
this we infer that,
{\it independent of any other measurement,
the first higgs mass measurement
will rule out one of the two main contenders for \underline{the} electroweak
theory: the SM, with no new physics below
$\sim 10^{10}$ GeV, or the MSSM with supersymmetry breaking scale
$M_{SUSY}\lsim 1$ TeV.}
Here we improve our previous calculation \cite{ourPRL} of
the renormalized MSSM higgs mass by including
two-loop QCD corrections and
then summing to all orders in perturbation theory the
leading and next--to--leading logarithms with a
renormalization group equation (RGE) technique \cite{marco,marcodpf}.
We also use the improved stability \cite{Casasetal,alta}
and metastability \cite{Quirosmeta}
lower bounds on the SM higgs mass (which we summarize in \S 2).

In the limit where the masses of the pseudoscalar, heavy and charged
Higgs bosons (these are $m_A$, $m_H$ and $m_{H^{\pm}}$, defined in \S
3) are large compared to $M_Z$ (of the order of a TeV for example),
the Feynman rules connecting the light
Higgs in the MSSM to ordinary matter are approximately equal to
the SM Feynman rules\cite{newHowie}. Therefore, in this limit,
the MSSM light Higgs looks very much like the SM Higgs in its
production channels and decay modes; the only difference,
a vestige of the underlying supersymmetry,
is that the constrained higgs self--coupling requires the MSSM higgs
to be light,
whereas SM vacuum stability requires the SM higgs to be heavy.  Thus,
{\it may only be possible to distinguish between the SM higgs and the
lightest higgs of MSSM (with $M_{SUSY}\lsim 1$ TeV)
by their allowed mass values.}
We demonstrate these allowed mass values in our Figures 1 and 2.
Furthermore, the mass of the lightest MSSM higgs rises toward its upper bound
as the ``other'' higgs masses are increased.
\footnote
{The saturation of the MSSM upper bound with increasing ``other'' higgs
masses
is well known in tree--level relations
(the bound $m_h \leq M_Z|\cos(2\beta)|$
approaches an equality as higgs masses increase) \cite{HHG}.
The MSSM upper bound still saturates
with increasing ``other'' higgs masses even when one--loop corrections
are included.
}
Thus, for masses in the region where the SM lower bound
and the MSSM upper bound overlap,
the SM higgs and the lightest MSSM higgs may not be distinguishable by
branching ratio or width measurements \cite{cpr1}.
Only if the two bounds are separated by a gap is this ambiguity avoided.

In the SM and even in supersymmetric models the main
uncertainty in radiative corrections is the value of the top mass.
With the announcement of the top quark mass,
this main uncertainty is greatly reduced.
{\it The radiatively corrected observable
most sensitive to the value
of the top mass is the mass of the
lightest higgs particle in susy models {\rm \cite{mt4}}}:
for large top mass, the top and scalar--top ($\tilde{t}$) loops dominate
all other loop corrections, and
{\it the light higgs mass-squared grows as }
$m_t^4 \ln(m_{\tilde{t}}/m_t)$.
\footnote
{It is not hard to understand this fourth power dependence; the contribution
of the top loop to the SM higgs self energy also scales as $m_t^4$.  However,
in the SM the higgs mass is a free parameter at tree--level, and so any
radiative correction to the SM higgs mass is not measurable.  In contrast,
in the MSSM the lightest higgs mass at tree--level is fixed by other
observables, and so the finite renormalization is measurable.
}
We quantify this large correction, including two-loop QCD corrections
and summing to all orders in perturbation theory the leading and
next-to-leading logarithms, in \S 3.

In addition to contrasting the MSSM with the SM, we also consider
in \S 4
supersymmetric models with a non-standard Higgs sector, in particular
the Minimal--plus--Singlet Susy
Standard Model [(M+1)SSM] containing an additional
SU(2) singlet, and the low energy effective theory of SUSY models
with a strongly interacting electroweak sector.
A discussion of supersymmetric grand unified theories (susy GUTs)
is put forth in \S 5; susy GUTs impose additional constraints on the
low energy MSSM, leading to a lower upper bound on the lightest higgs mass.
The discovery potential for the higgs boson is analyzed in \S 6, and
conclusions are presented in \S 7.

\section{Standard model vacuum stability bound}

It has been shown that when the newly reported
value of the top mass is input
into the renormalized effective potential for the SM higgs field, the
broken--symmetry potential minimum remains stable when the renormalization
scale
is taken all the way up to the Planck mass
only if the SM higgs mass satisfies the lower bound constraint \cite{alta}
\begin{equation}
m_H > 139+2.1(m_t-176)-4.5(\frac{\alpha_s - 0.118}{0.006}),
	\;\;\Lambda=10^{19}{\rm GeV}.
\label{eq:altarelli}
\end{equation}
In this equation,
mass units are in GeV, and $\alpha_s$ is the strong coupling constant
at the scale of the $Z$ mass. The accuracy of the bound is estimated
to be $\sim 5-10$ GeV.
A similar but slightly lower bound is found in ref. \cite{Casasetal}:
\begin{equation}
m_H > 136+1.92(m_t-176)-4.25(\frac{\alpha_s - 0.118}{0.006}),
	\;\;\Lambda=10^{19}{\rm GeV},
\label{eq:casasetal}
\end{equation}
valid in the range 150 GeV $< m_t <$ 200 GeV.
These equations are the result
of an analysis of the one-loop SM effective
potential using two loop beta functions and the
appropriate matching conditions.
Here the estimated accuracy is $\lsim 3$ GeV from the theoretical calculation,
and $\lsim 1$ GeV from the linear fit resulting in eq.\ (\ref{eq:casasetal}).

The definition of the SM which we use requires no new physics (\ie a desert)
``only'' up to the scale $\Lambda\sim 10^{10}$ GeV.
We use the $m_H$ vs.\ $m_t$ curves for various cut--off values
in ref. \cite{Casasetal}
to determine the coefficient of the $m_t$ term at $\Lambda\sim 10^{10}$ GeV;
and we run the SM renormalization group equations (RGE's)
to determine the coefficient of the $\alpha_s$ term at $\Lambda\sim 10^{10}$
GeV.
The resulting lower bound for $\Lambda\sim 10^{10}$ GeV is
\begin{equation}
m_H > 131+1.70(m_t-176)-3.47(\frac{\alpha_s - 0.118}{0.006}),
	\;\;\Lambda=10^{10}{\rm GeV}.
\label{eq:us}
\end{equation}
The accuracy of this bound should approximate that of
eq.\ (\ref{eq:casasetal}),
$\lsim 4$ GeV.
Because the parameter space for a smaller SM desert is necessarily contained
within
the parameter space for a larger SM desert, a smaller desert implies weaker
constraints
on the model; accordingly, we see that the lower bound on the higgs mass
relaxes
when the cut--off for new physics is reduced.  In fact,
it has been pointed out \cite{alta,Casasetal} that the discovery of
a higgs with low mass would place an upper limit on the scale of new physics.

This lower mass bound and the related ``triviality
bound''\cite{triv,lattice,wilsonrge}
are based on the physical requirement that the running higgs self--coupling
remains positive and finite up to the energy scale
$\Lambda$. Below this energy scale $\Lambda$ the
SM is supposed to be valid. If the higgs boson mass,
given by $\sqrt{2 \lambda}v_{SM}$,
is too small compared to the top quark mass, then
the running higgs self--coupling $\lambda$ turns
negative at a scale below the cut-off $\Lambda$\cite{lindner1}.
On the other hand, if the higgs boson mass is too
large, then the running self--coupling $\lambda$
diverges at a scale below the cut--off $\Lambda$.
Thus, for a given cut--off scale $\Lambda$
and top quark mass $m_t$,
the higgs mass is bounded from below by the
vacuum stability bound, and bounded from above
by the triviality bound.
For large values of the cut--off, $\Lambda\gsim 10^{10}$ GeV,
these bounds are only weakly dependent on the value of
$\Lambda$\cite{lindner2,Casasetal}.
By comparing eqn. (\ref{eq:us}) with eqns. (\ref{eq:altarelli}) and
(\ref{eq:casasetal}), we see that for a top quark mass $m_t=176$ GeV and
$\alpha_s = 0.118$, an increase in $\Lambda$ from $10^{10}$ GeV
to the Planck mass $\sim 10^{19}$ GeV raises the vacuum
stability bound by only 5 to 8 GeV.
To put it in simple terms:
if the running self--coupling $\lambda$ is going to
diverge or become negative, it will do so at a relatively low
energy scale.

It has been known for some time \cite{reviews}
that the SM lower bound
rises rapidly as the value of the top mass increases through $M_Z$;
below $M_Z$
the bound is of order of the Linde--Weinberg value, $\sim 7$ GeV \cite{LW}.
So what is new here
is the inference from the large reported value for $m_t$ that the SM higgs
lower mass bound dramatically exceeds 100 GeV!
Adding the statistical and systematic errors of the CDF top mass measurement
in quadrature gives a top mass
\footnote
{A top mass limit independent of the top decay modes is provided
by an analysis of the W boson width: $m_t > 62$ GeV at 95\%
confidence \cite{Wwidth}.
}
with a single estimated error of
$m_t=176\pm13$ GeV.
The D0 collaboration has also announced discovery of the top
quark\cite{D0}, with a top mass estimate of $199\pm 30$ GeV, consistent with
the (better--determined) CDF value.
The main uncertainty in the SM vacuum stability bound remains the exact value
of the top quark mass.
The CDF one--sigma uncertainty of 13 GeV in the top quark mass
translates into a 22 GeV one--sigma uncertainty in the bound of
Eq. (\ref{eq:us}).
The bound's dependence on the uncertainty in $\alpha_s$, a better known
parameter, is more mild.

It is possible that the observed vacuum state of our
universe is not absolutely stable,
but only metastable with a small probability to decay
via thermal fluctuations or quantum tunneling.
 If metastability rather than absolute stability is postulated,
then a similar but weaker bound results\cite{arnold}.
In an accurate  calculation of this metastability bound,
next--to--leading logs are included in the effective potential
and one--loop ring graph contributions to the Debye mass are
summed \cite{Quirosmeta}.

SM metastability bounds are given in ref.\ \cite{Quirosmeta} in tabular form
for $\alpha_s = 0.124$ and various values of $\Lambda$,
and in analytic form for $\Lambda =10^{19}$ GeV with various values of
$\alpha_s$.
To derive the metastability bound for our cut--off value $\Lambda=10^{10}$ GeV
and various $\alpha_s$ values, we do the following:
We first obtain the bound for $\alpha_s=0.124$ and $\Lambda=10^{10}$ GeV by
extrapolating
the values given in Table 1.\ of \cite{Quirosmeta}.  The
$\alpha_s$--dependent term at
$\Lambda=10^{19}$ GeV is obtained from eqn.\ (30) in \cite{Quirosmeta}.
Based upon our experience with running the SM RGEs from $\Lambda=10^{19}$
GeV down to
$\Lambda=10^{10}$ GeV for the SM stability bound, we note that the
coefficient of the
$\alpha_s$--dependent term is renormalized down by 20\%
(compare eqn. (\ref{eq:us})
to eqns. (\ref{eq:altarelli}) and (\ref{eq:casasetal})).  So we reduce the
coefficient
of $\alpha_s$ by 20\%. The change in the Higgs mass bound
effected by this renormalization is small, $\sim$ 1 or 2 GeV or less.
The resulting metastability bound at $\Lambda=10^{10}$ GeV is
\begin{equation}
m_H > 123+2.05(m_t-176)-3.9(\frac{\alpha_s - 0.118}{0.006})(\frac{m_t}{176}),
	\;\;\Lambda=10^{10}{\rm GeV}.
\label{eq:meta}
\end{equation}
According to eq.\ (\ref{eq:casasetal}), the linear fit is valid to better
than one GeV
for $m_h>60$ GeV, and the overall theoretical error is negligible compared
to the experimental errors in the $\alpha_s$ and $m_t$ values.

In our figures, we will present both the stability and
the metastability lower bounds.
The metastability bound is necessarily lower than the stability bound.
A comparison of eqns. (\ref{eq:us}) and (\ref{eq:meta}) shows that the
ordering is maintained in the $m_t$--region of interest, below 200 GeV;
beyond $m_t= 200$ GeV the fitted equations are no longer valid.
The CDF top mass values including $1\sigma$
allowances are 163, 176, and 189 GeV.
The vacuum stability bounds following from
Eq.~(\ref{eq:us})  for these top quark masses
with $\alpha_s = 0.118$ are 109, 131, and 153 GeV, respectively,
whereas the metastability bounds are 96, 123, and 150 GeV, respectively
\footnote
{LEP experiments have established
the \underline{non--existence}
of the SM higgs particle below a mass value of 64 GeV \cite{LEP}.
}.

As is evident in eqns.\ (\ref{eq:altarelli}), (\ref{eq:casasetal}),
(\ref{eq:us}),
and (\ref{eq:meta}),
the vacuum stability and metastability bounds on the SM higgs
mass are sensitive to the value of $\alpha_s(M_Z)$.
We have taken $\alpha_s=0.118$ (the central value in the work of \cite{alta})
to produce the bounds displayed in Fig. 1.
The 1994 world average derived by the Particle Data Group \cite{PDG94}
is $0.117\pm0.005$.
The value derived from fitting SM radiative corrections to LEP/SLC
precision electroweak data
is $\alpha_s(M_Z)=0.124\pm0.005$ in \cite{ellisetal}, and
$\alpha_s(M_Z)=0.122\pm0.005$ in \cite{SMpok}.
Other LEP analyses, and deep inelastic leptoproduction (Euclidean) data
extrapolated to
the $M_Z$ scale give lower values $\sim 0.112$;
a comparison of low $Q^2$ deep--inelastic data to the Bjorken sum
rule \cite{Bj}
yields \cite{E&K} $\alpha_s(M_Z)=0.116^{+0.004}_{-0.006}$.
The LEP working group \cite{who} quotes a world average
of $\alpha_s(M_Z)=0.120\pm0.006\pm0.002$, assuming the SM.
If we use  the generous
value $\alpha_s=0.130$, the stability bound on the SM higgs mass decreases
by about 9 GeV for $m_t>160$ GeV, and the metastability bound decreases
by about 8 GeV.

The vacuum stability bound on the SM higgs mass rises roughly linearly with
$m_t$, for $m_t \gsim 100$ GeV, whereas
the upper limit on the lightest MSSM higgs mass grows
quadratically with $m_t$.
Therefore, for very large values of the
top quark mass $m_t$, the two bounds will inevitably overlap.
In addition, for low values of $m_t$
the two bounds may overlap.
For example, for very large or very small values of  $\tan\beta$
the MSSM upper bound is at least $M_Z$, but the
SM lower bound is only 60 GeV for $m_t = 130$ GeV\cite{sher}.
However, for $m_t$ heavy, but not too heavy, there may be no overlap.
In what follows, we show that in fact for $m_t$ around the value reported
by the CDF collaboration, there is little ($\alpha_s=0.130$) or no
($\alpha_s=0.118$) overlap between the SM higgs mass
lower bound and the MSSM upper bound.
Thus, the first measurement of the lightest higgs mass
will probably suffice to exclude either the SM higgs sector,
or the MSSM higgs sector!

\section{The lightest higgs in the MSSM}

The spectrum of the higgs sector in the MSSM contains two CP--even
neutral higgses, $h$ and $H$, with $m_h<m_H$ by convention,
one CP--odd neutral higgs
$A$ and a charged higgs pair $H^{\pm}$. A common
convenience is to parameterize the higgs sector by the mass of the
CP--odd higgs $m_A$ and the vev ratio
$\tan\beta\equiv v_T/v_B$.
These two parameters completely
specify the masses of the higgs particles at tree level
\begin{eqnarray}
m_{H,h}^2= & \half(m_A^2+m_Z^2)\pm\half\sqrt{(m_A^2-m_Z^2)^2\cos^2 2\beta
+(m_A^2+m_Z^2)^2\sin^2 2\beta} \nonumber \\
m_{H^{\pm}}^2= & m_A^2+m_W^2
\label{eq:treeHiggs}
\end{eqnarray}
implying for example that $m_{H^{\pm}}>m_W$, that the upper bound
on the lightest higgs mass is given by
\beq
m_h\leq |\cos(2\beta)| \, M_Z,
\label{eq:tree}
\eeq
that the lightest higgs mass vanishes at tree level if $\tan\beta=1$,
and that the masses $m_H, m_A$, and $m_{H^{\pm}}$
all increase together as any one of them is increased.
However, radiative
corrections strongly modify the tree level predictions in the neutral
 \cite{mt4,neutral,DreesNojiri,diazhaberii} and
charged \cite{ChargedH,DreesNojiri,diazhaberi}
higgs sectors.  Some consequences are that the
charged higgs can be lighter than the $W$ gauge boson \cite{diazhaberi},
that the $\tan\beta=1$ scenario, in which $m_h=0$ at tree level,
is viable due to the possibility of a large radiatively
generated mass \cite{diazhaberii},
and that the upper bound on the
lightest higgs mass is increased by terms proportional to
$m_t^4 \ln(m_{\tilde{t}}/m_t)$,
as advertised in our introduction
\footnote{
Note that in the susy limit, $m_t=m_{\tilde{t}}$ and the fermion and boson
loop contributions cancel each other.  However, in the real world of broken
susy, $m_t\neq m_{\tilde{t}}$, and
the cancellation is incomplete.
The top quark gets its mass from its
yukawa coupling to the electroweak vev, whereas the scalar top mass arises
from
three sources, from D--terms, from the top yukawa coupling, but mainly from
the insertion into the model of dimensionful soft susy--breaking
parameters.  The interplay of these diverse masses
leads to the dramatic correction.
Note that the correction grows logarithmically
as $m_{\tilde{t}}$ gets heavy, rather than decoupling!
}
\cite{mt4}.

An important mechanism for the production of the neutral higgses
in $e^+e^-$ colliders is the brehmsstrahlung of a higgs by a $Z$ gauge boson.
Relative to the coupling of
the SM higgs to two $Z$ bosons, the $ZZH$ coupling is
$\cos(\beta-\alpha)$ and the $ZZh$ coupling is $\sin(\beta-\alpha)$,
where $\alpha$ is the mixing angle in the CP-even neutral higgs mass matrix.
The angle is restricted to $-\frac{\pi}{2}\leq \alpha \leq 0$, and is
given at tree level by
\begin{equation}
\tan2\alpha={{(m_A^2+m_Z^2)}\over{(m_A^2-m_Z^2)}}\tan2\beta.
\label{eq:tantwoalpha}
\end{equation}
 From Eq.(\ref{eq:tantwoalpha}) it is seen that
the limit $m_A\rightarrow\infty$ is important for three reasons. First,
it requires
$\alpha\rightarrow\beta-\pi/2$, implying that
$\cos(\beta-\alpha)\rightarrow 0$,
\ie, the heavy higgs decouples from the $Z$ gauge boson.
Secondly, it requires
that $\sin(\beta-\alpha)\rightarrow 1$, \ie, the light higgs behaves
like the SM higgs.
And thirdly, $m_A\rightarrow \infty$ is the limit in which
the tree level $m_h$ saturates its maximal value given in Eq. (\ref{eq:tree})
for any value of $\tan\beta$.

We use the diagramatic technique with an on-shell
renormalization scheme to calculate
the renormalized lightest MSSM higgs mass, $m_h$ \cite{marco}.  We include
the full one--loop corrections from the
top/bottom quarks and squarks, the leading--log corrections from
the remaining fields (charginos, neutralinos, gauge bosons, and higgs bosons)
\footnote{
Calculations of full one--loop corrections from all particles
\cite{1loop} have shown that finite (\ie non--logarithmic) corrections due
to loops
with particles other than the
top/bottom quarks and squarks are very small.
},
the dominant two--loop corrections,
and the full momentum--dependence of the higgs self--energies.
We then perform a Renormalization Group Equation (RGE) improvement
\cite{haberralf} of these results
in order to include the resummed leading and next--to--leading
logarithms.
The result is a highly accurate calculation of the lightest MSSM higgs mass,
perhaps the most accurate available in the literature.


We find the renormalized neutral higgs masses by looking for the zeros
of the determinant of the inverse propagator matrix, including
the loop corrections \cite{marcodpf}.  The two solutions to
\begin{equation}
\Sigma^{\chi}_{11}(p^2)\Sigma^{\chi}_{22}(p^2)=
\left[\Sigma^{\chi}_{12}(p^2)\right]^2,
\label{eq:detinvprop}
\end{equation}
are the pole higgs masses $p^2=m_h^2$ and $p^2=m_H^2$.
The propagators are calculated in a basis in which the
CP-even higgs fields $\chi_1$ and $\chi_2$ are unmixed at tree-level.
We renormalize
each matrix element of the inverse propagator matrix
first, and later diagonalize it nonperturbatively.
Furthermore, we keep the full momentum dependence of the self energies
in eq. (\ref{eq:detinvprop}). This is equivalent to defining a
momentum dependent mixing angle $\alpha(p^2)$.  With this procedure, we
avoid the introduction of a mixing angle counterterm, which allows us to
calculate directly the renormalized mixing angle at the two physically
relevant scales $\alpha(m_h^2)$ and $\alpha(m_H^2)$ \cite{marcodpf}.

Two--loop corrections are negative and decrease the upper bound
of the higgs mass by several GeV \cite{2loop}.
We include the
dominant two--loop corrections of ref. \cite{2loop} which
include the leading and next-to-leading logarithms.
Finally, using an RGE technique,
we extend the results of ref.\cite{2loop} by
summing to all orders in perturbation
theory these leading and next--to--leading logarithm terms.
In order to do this, we solve the two-loop RGE \cite{eq} with a supersymmetric
boundary condition at the scale $M_{SUSY}$ to obtain the
quartic higgs self--coupling constant at the weak scale.
In this way, the running higgs mass squared is equal to $\lambda v^2$,
where $v^2=v_T^2+v_B^2$ ($v_T$ and $v_B$ are the vevs
of the two higgs doublets.).
This RGE improvement\cite{diazhaberii,2loop},
\begin{equation}
(\Delta m_h^2)_{RGE}=\lambda v^2-M_Z^2\cos^2 2\beta -
(\Delta m_h^2)_{lnll}.
\label{eq:RGEimprov}
\end{equation}
depends of course on the value of the top quark mass.
Here $(\Delta m_h^2)_{lnll}$ contains the logarithmic part of the
one-- and two--loop corrections, the so--called leading and next--to--leading
logarithms.
For example, at $m_t\sim 176$ GeV, we find the RGE correction to be --2
to --3
GeV for large $\tan\beta$ and --5 to --7 GeV if $\tan\beta$ is small.
We include this correction in all of our plots.

We choose $m_A$ and all squark mass parameters to be large, equal to
1 TeV
\footnote{
We note that $\lsim 1$ TeV emerges naturally for the heavier superparticle
masses when the MSSM is embedded into a GUT \cite{llabf,kkrw,lnpwz}.
},
in order to find the maximum light higgs mass.
With respect to the squark mixing, we work in three extreme scenarios:\\
(a) no mixing,
\ie, $\mu=A_t=A_b=0$, where $\mu$ is the supersymmetric higgs mass
parameter and $A_i$, $i=t,b$ are the trilinear soft supersymmetry breaking
terms; and \\
maximal mixing\\
(b) with $\mu=A_t=A_b=1$ TeV, \\
(c) and $\mu=-1$ TeV, $A_t=A_b=1$ TeV.\\
We mention again that our chosen definition for the MSSM is the
conventional one,
with  $M_{SUSY}$, all of the soft supersymmetry breaking
terms, and  $\mu$, having a magnitude of at most 1 TeV. One of the
motivations for this choice is that in supergravity
models the electroweak symmetry can be broken radiatively
without fine--tuning the initial parameters, if $M_{SUSY}$ is not
too large\cite{BdCyC}.

The resulting lightest higgs mass as a
function of $\tan\beta$ is shown in Fig. 1 for the
CDF central value of the top quark mass and the $\pm 1 \sigma$ mass values.
The accuracy of this bound can be estimated to be $\sim 10$ GeV, which
is the difference between the one-- and two--loop bound calculated before
the RGE--resummation.
For the case $\tan\beta\sim 1$, the SM lower bound and the MSSM upper bound
are separated already at $m_t=163$ GeV\@.
Were if not for the SM \ul{meta}stability lower bound,
the gap would exist for all values of $\tan\beta$.
However, with the SM metastability bound,
it is not until $m_t \sim 175$ GeV that
a gap exists for all values of  $\tan \beta$.
In particular, for the preferred CDF value of $m_t=176$ GeV,
the two bounds do not overlap, making it
possible to distinguish the SM and the MSSM solely on
the basis of a determination of the higgs mass.
Even for $m_t=189$ GeV the gap is still increasing
with increasing top mass,
indicating that the eventual closing of the gap occurs at
still higher values of $m_t$.

Should $\alpha_s$ turn out to be closer to 0.130 than to the value 0.118
assumed here, then the separation of the SM higgs mass region
from the MSSM higgs mass region is not quite complete. We have seen that
the stability and metastability lower bounds on the SM higgs mass
decrease as $\alpha_s$ is increased.
The MSSM mass upper bound also decreases with increasing $\alpha_s$, but
at a much smaller rate. We find that raising $\alpha_s$ from $0.118$
to $0.130$
shifts the MSSM higgs mass bound by
$-0.5$ GeV for $m_t=$163 and by $-0.8$ GeV for $m_t=$189 GeV.
The result is that the gap apparent for all values of $\tan\beta$ in the
$\alpha_s=0.118$, $m_t=176$ GeV case (displayed in our Fig. 1b),
remains a gap in the $\alpha_s=0.130$ case only in the $\tan\beta\sim$
1 to 2 region.  However, the overlapping mass region for the remaining
$\tan\beta$ values is small.  The region of overlap is interesting
only if the observed higgs mass turns out to lie in this region.
With a small overlap region, such an occurence is {\it a priori} unlikely.
A further (interesting) complication is that the best fit value for
$\alpha_s$, when
MSSM radiative corrections are assumed and fitted to precision data, is
\cite{MSSMpok}
$\alpha_s (M_Z) = 0.114\pm 0.007$.  This lower value suggests that it may be
best to
compare SM bounds with a given value assumed for $\alpha_s$ to MSSM bounds
with a
slightly lower value assumed for $\alpha_s$.

In Fig. 1 we can see that scenario (c) gives us a significantly
larger range of higgs mass values close to $\tan\beta\sim1$. This can be
understood in the $\tan\beta=1$ approximation:
there are non-leading logarithmic contributions to the higgs mass from loops
involving the top quark and squarks that are proportional to
powers of $(\mu-A_i)/m_t$ \cite{diazhaberii}.
Also in Fig. 1 we see that scenarios (b) and (c) offers a larger
value for the $m_h$ maximum than does scenario (a), except for
the region $\tan\beta\gg 1$. The reason is that among the additional
light higgs mass terms in (b) is a negative term proportional
to $-(\mu m_b/\cos\beta)^4$,
which becomes large \cite{haberralf}
when $\tan\beta\gg 1$.
More significant is the fact that the extreme values in (a), (b) and (c)
yield a very similar absolute upper bound in the region of acceptable
$\tan\beta$ values, thereby suggesting insensitivity of the MSSM upper
bound to a considerable range of the squark mixing parameters.

In the literature there are three popular methods to calculate the
renormalized higgs mass.
These are the RGE technique, the effective potential method,
and the diagramatic technique.
It is informative to compare these techniques, and to point out the
advantages of the approach we have undertaken.
The RGE technique is used for example in ref. \cite{eq},
where the leading and next-to-leading logarithms are summed to all orders
in perturbation theory to give
the running higgs mass. This technique is based on the
fact that the Veltman functions \cite{Veltman} which appear in
the diagrammatic method can be approximated by logarithms when
there are two different scales in the problem. The RGE technique
sums these logarithms to all orders, but drops all
non--logarithmic, finite terms.  These terms are often very important
\cite{diazhaberii,diazhaberi}.  Moreover, the reliability of the RGE
treatment
of the logarithmic terms decreases if the two scales are not very far
apart (as is the case here, where the two scales are the EW and SUSY
breaking scales).
Numerically, the higgs mass calculated with the RGE method
can differ by 10 GeV or more compared to the diagramatic method,
even if two--loop RGEs are used.

The renormalization group improvement (see our
eq. (\ref{eq:RGEimprov})) we use in our work
replaces the logarithmic part of the corrections
obtained with the diagrammatic method by the resummed logarithmic
corrections
as obtained with the renormalization group technique. Our results
therefore incorporate both the important finite corrections at the
two-loop level and the
resummed leading and next--to--leading logarithmic  corrections.

The second popular technique is the effective potential method.
In ref.
\cite{diazhaberii} the effective potential method is compared with the
diagramatic technique. Working in an on--shell scheme in both
methods, it is shown that the two techniques reproduce the same
answer when the tree level higgs mass is zero and when all
supersymmetric particles are included in the effective
potential. On the contrary,
if the tree level higgs mass is non--zero, the effective potential
answer has to be corrected using diagramatic methods.
With these diagrammatic corrections,
the two methods become indistinguishable.

The effective potential method is used in ref. \cite{CEQR}.
There the $\overline{MS}$ renormalization scheme is also used
and so the comparison with our on--shell
diagramatic method is not simple.
A non--trivial ambiguity for the choice of the arbitrary scale
is present in this method.
A further limitation in this calculation is the inclusion in the effective
potential
of only SM particles. Important log terms arising from
susy particle loops are therefore absent.  When the susy particles
are ignored, the only connection with supersymmetry
is in the boundary condition for $\lambda$ at the scale $M_{SUSY}$.
A partial compensation is made
by including the threshold effects of susy particles in the
form of step--functions.
What would be a full Veltman's function in the diagramatic method
is approximated in the effective potential method
by a step function shift\cite{ralfthresh} in
the boundary condition:
$\lambda={\textstyle{1\over 4}}(g^2+g'^2)\cos^2 2\beta+\Delta\lambda$.

In our diagramatic method these approximations are not present since
the effects of the non--logarithmic terms are included in the full
expressions of the Veltman's functions. For example,
important non--logarithmic effects are included, such as
the decreasing of the higgs mass when
$\tan\beta\rightarrow\infty$, $\mu\sim 1$ TeV and $A=0$, as
explained above and seen in Fig. 1.
Also, the effect of large splitting in the
masses of the stop squarks is automatically
taken into account in our diagramatic method.
These effects are not included in ref. \cite{CEQR}.

There are two further improvements that we have achieved.
The first improvement is the use of different RGEs above and below the top
quark mass.  Below $m_t$ the top quark mass decouples and the RGE for
$\lambda$
does not contain the top Yukawa coupling.
This effect can be important.
In principle, the RGE for the gauge couplings should also be modified.
In practice, it is a negligible effect.
(This modification is more complicated, since the electroweak
gauge symmetry is broken. A careful analysis can be found in
ref. \cite{haberralf}.)
The second improvement is the consideration of the running of $\tan\beta$.
In practice, this effect is numerically small\cite{haberralf}.

We finish this section with some comments on the decay
$b\rightarrow s\gamma$.
It is known that the branching ratio $B(b\rightarrow s\gamma)$ has
a strong dependence on the susy higgs
parameters \cite{bsfSUSY,BdCyCbsf,diazbsfi}.
However, when all squarks are heavy, as here, the contribution
from the chargino/squark loops to $B(b\rightarrow s\gamma)$
is suppressed. In the case of heavy squarks, the
charged--higgs/top--quark loop may seriously alter the rate, and
strong constraints on the charged higgs minimum
mass result \cite{hewettBBP,diazbsfi}.
This constraint does not affect the present work,
where we take $m_A$ and therefore $m_{H^{\pm}}$ and $m_H$ large
in order to establish the light higgs upper bound:
in the large $m_A$, large squark mass limit,
the ratio $B(b\rightarrow s\gamma)$ approaches the SM value,
consistent with the CLEO bound \cite{expbsf}.

\section{The lightest higgs in non-standard susy models}

The MSSM can be extended in a
straightforward fashion by adding an $SU(2)$ singlet $S$ with vanishing
hypercharge to the theory \cite{Ell}.
As a consequence, the particle spectrum contains an
additional scalar, pseudoscalar, and neutralino.
This extended model, the so--called (M+1)SSM, features four possible
additional terms in the superpotential.
Two of these terms,  $\lambda S H_B \epsilon H_T$ and
$\frac{1}{3}\kappa S^3$,
enter into the calculation of the lightest higgs mass;
$\lambda$ enters directly, while $\kappa$ enters through the RG equations.
$\epsilon$ is the usual antisymmetric 2 by 2 matrix.

At tree-level, a study of the eigenvalues of the scalar mass matrix
gives an upperbound on the mass of the lightest higgs boson:
\begin{equation}
m_h^2 \leq M_Z^2 \left\{ \cos^2 2 \beta +2 \frac{\lambda^2}{g_1^2 + g_2^2}
\sin^2 2 \beta \right\}.
\end{equation}
The first term on the right hand side
is just the MSSM result of Eq. (\ref{eq:tree}).
The second term gives a positive contribution,
and since the parameter $\lambda$ is {\it a priori} free,
weakens the upperbound considerably \cite{Vel,Kan}.
However, there are two
scenarios in which the bound prooves to be very restrictive.
In the first scenario $\tan\beta$ is large, and therefore $\cos^2 2\beta$ is
necessarily $\gg \sin^2 2\beta$.  In the second scenario
the value of $\lambda$ is limited by the assumption of
perturbative unification. In this latter scenario, even if
$\lambda$ assumes a high value at the GUT scale, the renormalization
group equations drive the evolving value of $\lambda$ to
a moderate value at the SUSY breaking scale.
The exact higgs mass upper bound depends on the value
of the top yukawa $g_t$ at the GUT scale through the renormalization group
equations.  Above $M_{SUSY}$ the running of the coupling
constants is described by the (M+1)SSM renormalization group
equations, whereas below this scale the SM renormalization group equations
are valid. At $M_{SUSY}$ the boundary conditions
\begin{eqnarray}
\lambda^{SM} & = & \frac{1}{8} \left( g_1^2 +g_2^2 \right)
\left( \cos^2 2 \beta + 2 \frac{\lambda^2}
{ g_1^2 + g_2^2 } \sin^2 2 \beta \right),
\nonumber \\
g_t^{SM} & = & g_t \sin \beta,
\label{eq:rge}
\end{eqnarray}
incorporate the transition from the (M+1)SSM to the SM. Here
$\lambda^{SM}$ and $g_t^{SM}$ are the standard model higgs self--coupling
and top quark yukawa coupling respectively. The value of the
higgs boson mass is determined implicitly by the equation
$2 \lambda^{SM} \left( m_h \right) v_{SM}^2 = m_h^2$.
This RGE procedure of running couplings from $M_{SUSY}$ down to the weak
scale
takes into account logarithmic radiative corrections to the higgs boson mass,
including in particular those caused by the heavy top quark.

In Fig. 2 we show the maximum value of the
higgs boson mass as a function of $\tan \beta$ for
the chosen values of the top quark mass $m_t$.
We have adopted a susy--breaking
scale of $M_{SUSY}=1\ TeV$; this value is consistent with the notion of
stabilizing  the weak--to--susy GUT hierarchy, and is the value favored by
RGE
analyses of the observables $\sin^2\theta_W$ and $m_b/m_{\tau}$.
The bounds in Fig. 2 are quite insensitive to the choice of $M_{SUSY}$,
increasing very slowly
as $M_{SUSY}$ increases \cite{Vel}.
We have assumed that all superpartners and all higgs bosons except
for the lightest one are heavy, \ie $\sim M_{SUSY}$.
For low values of the top quark mass ($\sim M_Z$), the mass upper bound on
the
higgs boson in the
(M+1)SSM will be substantially higher than in the MSSM at
$\tan\beta \lsim$ a few. This is because $\lambda(m_h)$ is large for low
$m_t$,
and because $\sin^2 2\beta \gsim \cos^2 2\beta$ for $\tan\beta \lsim$ a few.
However, for a larger top quark mass, as in Fig. 2, the difference
between the MSSM and (M+1)SSM upper bounds diminishes.
This is because $\lambda(m_h)$ falls with increasing $m_t$, and because
there is an increasing minimum value for
$\sin\beta=g_t^{SM}/g_t$ [from the second of Eqs. (\ref{eq:rge})],
and therefore for $\tan\beta$,
when $m_t\propto g_t^{SM}$ is raised and $g_t$ is held to be perturbatively
small up to the GUT scale\footnote
{Keeping $g_t$ perturbatively small up to the GUT scale implies
$m_t\leq$ its pseudo fixed--point value of $\sim 200 \sin\beta$.
Therefore, a measured top mass as large as that reported by CDF
requires $\tan\beta > 1$ in the GUT scenario,
and suggests saturation of the fixed--point.
}.
This increasing minimum value of $\tan\beta$ is evident in the curves of
Fig. 2.
A comparison of Figs. 1 and 2 reveals that the (M+1)SSM and MSSM bounds are
very similar at $\tan\beta\gsim 6$.
For $m_t$ at or above the CDF value, only this  $\tan\beta\gsim 6$
region is viable in the (M+1)SSM model.

In a fashion very similar to the (M+1)SSM, perturbative unificaton
yields a bound on the mass of the lightest higgs bosons in
more complicated extensions of the MSSM.
In general, the lowest eigenvalue of the scalar mass matrix
is bounded by $M_Z$ times a factor which
depends on the dimensionless coupling constants in the
higgs sector. The renormalizaton group equations force these
coupling constants to assume relatively low
values at the SUSY breaking
scale, and as a consequence the mass bound on the
lightest higgs boson is of the order of $M_Z$.

Although a bound on the mass of the lightest higgs boson
exists in perturbative SUSY models, this is not the
case in SUSY models with a strongly interacting symmetry
breaking sector. The low energy physics of this
class of theories is described by a supersymmetric non-linear
sigma model, which is
obtained by imposing the constraint $H_T \epsilon H_B = \frac{1}{4} v_{SM}^2
\sin^2 2 \beta$ on the action of the MSSM \cite{Fer}.
This constraint is the only one possible in the MSSM higgs sector that
obeys supersymmetry, is invariant under $SU(2)\times U(1)$,
and leaves the vev in a global minimum
\footnote
{This MSSM non--linear sigma model is not the formal heavy higgs limit of
the MSSM, but is a heavy higgs limit of the (M+1)SSM; the MSSM does
not contain an independent, dimensionless, quartic
coupling constant $\lambda$ in the higgs sector which can be taken to
infinity,
whereas the (M+1)SSM (and the SM) does.
}.
As a result of this
constraint one of the scalar higgs bosons, the pseudoscalar, and one of
the neutralinos are eliminated from the particle spectrum. The remaining
higgs
boson has a mass
$m_h^2 =  M_Z^2 + (\hat{m}_T^2 + \hat{m}_B^2) \sin^2 2 \beta$, and the
charged
higgs bosons have masses
$m_{H^{\pm}}^2 = M_W^2 + (\hat{m}_T^2 + \hat{m}_B^2)$.
Here, $\hat{m}_T^2$ and $\hat{m}_B^2$
are soft, dimensionful, susy--breaking terms; they may be positive or
negative.

In order for the
notion of a supersymmetric non-linear model to be relevant, the susy
breaking scale is required to be much smaller than the chiral
symmetry breaking scale $4 \pi v_{SM}$. The natural magnitude
for the parameters
$\hat{m}_B^2$ and $\hat{m}_T^2$
is therefore of the order of $M_Z^2$. Consequently,
both the neutral and the charged higgs bosons have masses of at most a
few multiples of $M_Z$ in the non--linear model.
This formalism of the effective action allows a description of the low
energy
physics {\it independent} of the particular strongly--interacting
underlying theory from which it derives.
Thus we believe that the non--linear MSSM model presented here is probably
representative of a class of underlying strongly--interacting susy models.
The lesson learned then is that measuring a
value for $m_h$ at $\lsim 300$ GeV cannot validate the SM, MSSM, (M+1)SSM,
or any other electroweak model.  However, the premise of this present
article
remains valid, that
such a measurement should rule out one or more of these popular models.

\section{Supersymmetric Grand Unified Theories}

Supersymmetric grand--unified theories (susy GUTs)
are the only simple models in which the three low energy gauge
coupling constants are known to merge at the GUT scale,
and hierarchy and parameter--naturalness issues are solved.
Thus, it is well motivated to consider the grand unification of the low
energy susy models.
At low energies, SUSY GUT models reduce to the
MSSM, but there are additional relations between the parameters \cite{kkrw}.
The additional constraints must yield an effective low energy theory
that is a special case of the general MSSM we have just considered.
Therefore, the upper bound
\footnote
{In fact, the additional restrictions may be so constraining as to also
yield a {\it lower} limit on the lightest higgs mass,
in addition to the upper limit.
For example, $m_h> 85$ GeV for $\tan\beta > 5$ and $m_t=170$ GeV is
reported in
ref.\cite{kkrw}, and a similar result is given in \cite{lnpwz}.
}
on $m_h$ in such SUSY GUT models is in general {\it lower}
than in the MSSM without any restrictions.
The assumption of gauge coupling constant unification
(with its implied desert between $M_{SUSY}$ and $M_{GUT}$) presents no
significant constraints on the low energy MSSM parameters \cite{kkrw,lang1}.
However, the further assumption that the
top yukawa coupling remains perturbatively small up to $M_{GUT}$
leads to the low energy constraint $0.96 \leq \tan\beta$.  This is because
the RGE evolves a large but perturbative top yukawa coupling
at $M_{GUT}$ down to its well--known infrared pseudo--fixed--point
value at $M_{SUSY}$ and below, resulting in the top mass value
$\sim 200 \sin\beta$ GeV.
If the bottom yukawa is also required to remain perturbatively small
up to $M_{GUT}$, then $\tan\beta \leq 52$ \cite{bbk} emerges as a second
low energy constraint.

The pseudo--fixed--point solution is not a true fixed--point,
but rather is the low energy yukawa value that runs to become a Landau pole
(an extrapolated singularity, presumably tamed by new physics)
near the GUT scale.
The apparent CDF top mass value is within the estimated range of the
pseudo--fixed--point value.  Thus it is attractive to assume
the pseudo--fixed--point solution.
With the additional assumptions
that the electroweak symmetry is radiatively broken \cite{drees}
(for which the magnitude of the top mass is crucial) and
that the low energy MSSM spectrum is defined by a small number of
parameters
at the GUT scale (the susy higgs mass parameter $\mu$ -- which is also
the higgsino mass, and four universal soft susy--breaking mass
parameters: the scalar mass, the bilinear and trilinear masses, and the
gaugino mass),
two compact, disparate ranges for $\tan\beta$ emerge:
$1.0\leq \tan\beta \leq 1.4$ \cite{bbk},
and a large $\tan\beta$ solution $\sim m_t/m_b$.
\footnote
{It may be noteworthy that a fit of MSSM radiative corrections to the
electroweak
datum $R_b\equiv \Gamma (Z\rightarrow b{\bar b}) / \Gamma
(Z\rightarrow hadrons)$
reveals a preference for just these two $\tan\beta$ regions \cite{MSSMpok}.
}
Reference to our Figs. 1 and 2 shows that the gap between the SM and MSSM
is maximized in the small $\tan\beta$ region and minimized in the large
$\tan\beta$ region, whereas just the opposite is true for the gap between the
SM and (M+1)SSM models.  Moreover, the (M+1)SSM model is an inconsistent
theory
in the small $\tan\beta$ region if $m_t \gsim$ 160 GeV.

In fact, a highly constrained
low $\tan\beta$ region $\sim 1$ and high $\tan\beta$ region
$\gsim$ 40--70
also emerge when
bottom--$\tau$ yukawa unification at the GUT scale is imposed on
the radiatively broken model \cite{fixpt,barger,lang2,copw}.
Bottom--$\tau$ yukawa coupling unification is attractive in that it is
natural in susy SU(5), SO(10), and $E_6$,
and explains the low energy relation, $m_b\sim 3 m_{\tau}$.
With bottom-$\tau$ unification,
the low to moderate $\tan\beta$ region requires
the proximity of the top mass to its fixed--point value \cite{bcpw},
while the high $\tan\beta$ region also requires the proximity of the bottom
and $\tau$ yukawas to their fixed--point; the emergence of the two
$\tan\beta$
regions results from these two possible ways of assigning fixed--points.

The net effect of the yukawa--unification constraint in susy GUTs is
necessarily to widen the mass gap
between the light higgs MSSM and the heavier higgs SM,
thus strengthening the potential for experiment to distinguish the models.
The large $\tan\beta$ region is
disfavored by proton stability \cite{arnonath}.
Adoption of the favored low to moderate $\tan\beta$ region
leads to a highly
predictive framework for the higgs and susy particle
spectrum \cite{lang2,copw}.
In particular, the fixed--point relation $\sin\beta \sim m_t/(200 GeV)$
fixes $\tan\beta$ as a function of $m_t$.
For a heavy top mass as reported by CDF, one has $\tan\beta \sim$ (1, 2)
for $m_t =$ (140, 180) GeV.
Since $\tan\beta \sim 1$ is the value for which
the $m_h$ upper bound is minimized (the tree--level contribution to $m_h$
vanishes),
the top yukawa fixed--point models offer a high likelihood for $h^0$
detection at LEP200.
Reduced $m_h$ upper bounds have been reported in \cite{barger,lang2}.
The reduction in these bounds is due to the
small $\tan\beta$ restriction, an inevitable consequence of assigning the
top mass, but not the bottom mass, to the pseudo fixed--point.
These bounds are basically our bound
in Fig. 1 for $\tan\beta\sim 1$, when allowance is made for small
differences
resulting from different methods and approximations.

Even more restrictive susy GUTs have been analyzed.
These include the ``no--scale'' or minimal supergravity models \cite{gp},
in which the soft mass parameters $m_0$ (universal scalar mass)
and $A$ are zero at the GUT
scale; and its near relative, the superstring GUT,
in which the dilaton vev provides the dominant source of susy
breaking and so $m_0$, $A$, and the gaugino mass parameter all scale
together at the GUT scale \cite{bgkp}.
Each additional constraint serves to further widen the SM/MSSM higgs
mass gap.

In radiatively broken susy GUTs with universal
soft parameters, the superparticle spectrum emerges at $\lsim$ 1 TeV.
If the spectrum in fact saturates the 1 TeV value, then as we have seen
the Feynman rules connecting $h^0$ to SM particles
are indistinguishable from the Feynman rules of the SM higgs.
Thus, it appears that
if a susy GUT is the choice of Nature, then the mass of the lightest higgs,
but not the higgs production rate or dominant higgs decay modes,
may provide our first hint of grand unification.

\section{Discovery potential for the higgs boson}

The higgs discovery potential of LEPII \cite{gunion,djouadi}
depends on the energy at
which the machine is run.  A SM higgs mass
up to 105 GeV is detectable at LEPII with the $\sqrt{s}=200$ GeV option
(LEP200), while a SM higgs mass only up to 80 GeV is detectable with LEP178.
As we have shown, with
the large value of $m_t$ reported by CDF, the upper limit
on the MSSM $h^0$ mass is $\sim 120$ GeV. This limit is $\sim 10$ GeV
lower than that reported in our previous work\cite{ourPRL},
as a result of the inclusion of
RGE--resummed leading and next-to leading logarithms and
2--loop finite QCD corrections.
Near this upper limit the MSSM higgs has the production and decay
properties of the SM higgs.  Discovery of this
lightest MSSM higgs then argues strongly for the LEP200 option over LEP178.
Furthermore, for any choices of the MSSM parameters,
associated production of either $h^0 Z$ or $h^0 A$ is guaranteed
at LEP200 as long as $m_{\tilde{t}}\lsim$ 300 GeV \cite{gunion}.
Even better would be LEP230, where detection of $Z h^0$ is guaranteed
as long as $m_{\tilde{t}}\lsim$ 1 TeV \cite{gunion}.
At an NLC300 (the Next Linear Collider), detection of $Z h^0$ is guaranteed
for MSSM or for (M+1)SSM \cite{gunion}.
Turning to hadron colliders \cite{zerwas,mrenna}, it is
now believed that while the SM higgs cannot be discovered
at Fermilab's Tevatron with its present energy and luminosity,
the mass range 80 GeV to 130 GeV is detectable at any hadron collider
with $\sqrt{s} \gsim 2$ TeV
and an integrated luminosity $\int dt\,{\cal L} \gsim 10 {\rm fb}^{-1}$
\cite{mrenna};
the observable mass window widens significantly with increasing luminosity,
but very little with increasing energy.
For brevity, we will refer to this High Luminosity DiTevatron hadron machine
as the ``HLDT''.
If the SM desert ends not too far above the electroweak scale, then the
SM higgs may be as heavy as $\sim$ 600--800 GeV
\footnote
{Theorists would prefer an even lower value of $\lsim 400$ GeV,
so that perturbative calculations in the SM converge \cite{durand}.
}
(but not heavier, according to the triviality argument),
in which case only the LHC (and not even the NLC500) guarantees detection.

We present our conclusions on detectability for
the CDF central $m_t$ value, for the $m_t \pm 1\sigma $ values,
and for a $m_t-3\sigma$ value of 137 GeV:\\
(i) if $m_t \sim 137$ GeV,
the SM higgs mass lower bound from absolute vacuum stability
is equal to the experimental lower bound
of $m_H=64$ GeV,
while the metastability bound allows a mass as low as 43 GeV
\footnote
{Recall that for the SM vacuum stability and metastability
bounds we assume a desert up to $\sim 10^{10}$ GeV.
};
a SM mass up to (80, 105, 130) GeV is
detectable at (LEP178, LEP200, HLDT);
and the MSSM $h^0$ is
certainly detectable at LEP178 for $\tan\beta \sim$ 1--2, and
certainly detectable at LEP200 for all $\tan\beta$.\\
(ii) if $m_t \sim 163$ GeV, then
the abolute (metastability) SM lower bound rises to 109 (96) GeV,
so the SM higgs cannot be detected at LEP178 and probably not at LEP200,
but is still detectable at the HLDT if its mass is below 130 GeV;
the lightest MSSM higgs is certainly detectable at LEP178 if $\tan\beta$ is
very close to 1, and is certainly detectable at LEP200 if $\tan\beta$
is $\lsim 3$.\\
(iii) if $m_t \sim 176$ GeV, then
the SM higgs is above 131 (123) GeV, out of reach for LEPII and probably
the HLDT
as well;
the MSSM higgs is certainly detectable at LEP200 if $\tan\beta \sim$ 1--2.\\
(iv) if $m_t \sim 189$ GeV, then
the SM higgs is above 153 (150) GeV in mass;
at any $\tan\beta$ value, the MSSM higgs is not guaranteed to be
detectable at LEP200,
but is certainly detectable at the HLDT if $\tan\beta \sim$ 1--3.\\
For these mass bounds the value $\alpha_s = 0.118$ has been assumed.
The MSSM mass upper bound is relatively insensitive to changes in $\alpha_s$,
whereas the SM mass lower bounds decrease about 3 GeV with each 0.005
increase in $\alpha_s$.
It is interesting that the $h^0$ mass range is most accessible to experiment
is $\tan\beta \sim$ 1--3, just the parameter range favored by susy GUTs.

\section{Discussion and conclusions}

For a top quark mass $\sim 176$ GeV, the central value reported by CDF,
and an $\alpha_s$ value of $\sim 0.118$,
a measurement of the mass of the higgs boson will distinguish
the SM with a $\gsim 10^{10}$ GeV desert from the MSSM with a
SUSY breaking scale of about $1$ TeV.
For the (M+1)SSM with the assumption of perturbative unification,
conclusions are similar to those of the MSSM.
For $\alpha_s$ above $0.120$ and $m_t \sim 176$ GeV,
a small overlap of the SM and MSSM mass regions exists,
but it is a priori
unlikely that the higgs mass will be found in this small range.
Accordingly, the first higgs mass measurement
can be expected to eliminate one of these popular models.

Most of the range of the lightest MSSM higgs mass is accessible to LEPII.
The lightest MSSM higgs is guaranteed detectable at LEP230 and at the LHC;
and the lightest (M+1)SSM higgs is
guaranteed detectable at a NLC300 and at the LHC.
Since there is no lower bound on the lightest MSSM higgs mass other than the
experimental bound,
the MSSM $h^0$ is possibly detectable even at LEP178 for all $\tan\beta$,
but there is no guarantee.
In contrast, the SM higgs is guaranteed detectable only at the LHC;
if $m_t \sim 176$ GeV, then according to the vacuum stability (metastability)
argument,
the SM higgs mass exceeds 131 (123) GeV, and so likely
will not be produced until the LHC or NLC is available.

Thus, one simple conclusion is that LEPII has a tremendous potential to
distinguish MSSM and (M+1)SSM symmetry breaking from SM symmetry breaking.
If a higgs is discovered at LEPII, the higgs sector of the
SM with a large desert is ruled out.
\vspace{1.0cm}
\\
{\bf Acknowledgements:}\\
This work was supported in part by the U.S. Department of Energy grant
no.\ DE-FG05-85ER40226, and the Texas National Research Laboratory Commission
grant no.\ RGFY93--303.
\vfill\eject

\vfill\eject
%
%
%
{\bf Figure Captions:}\\
\noindent
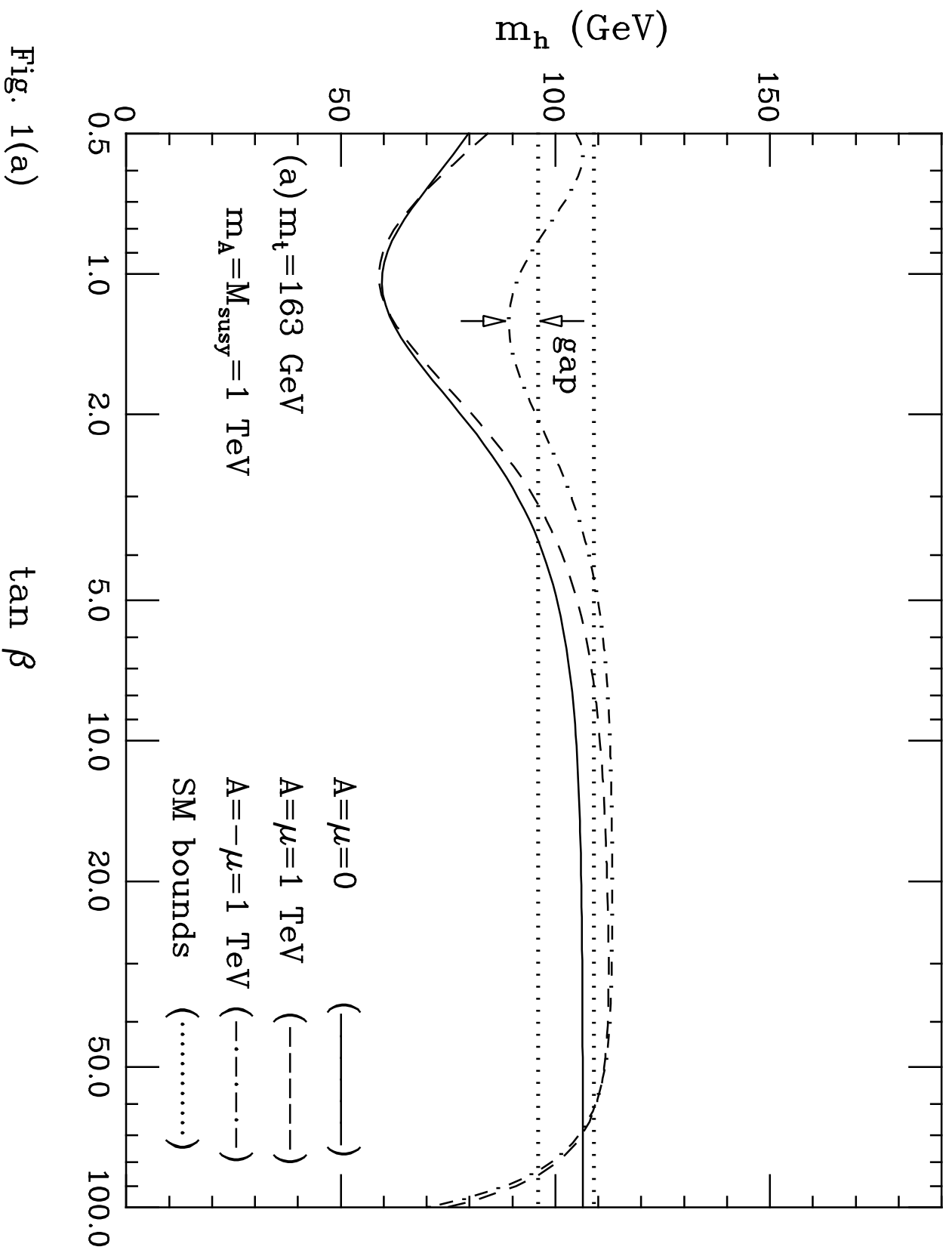
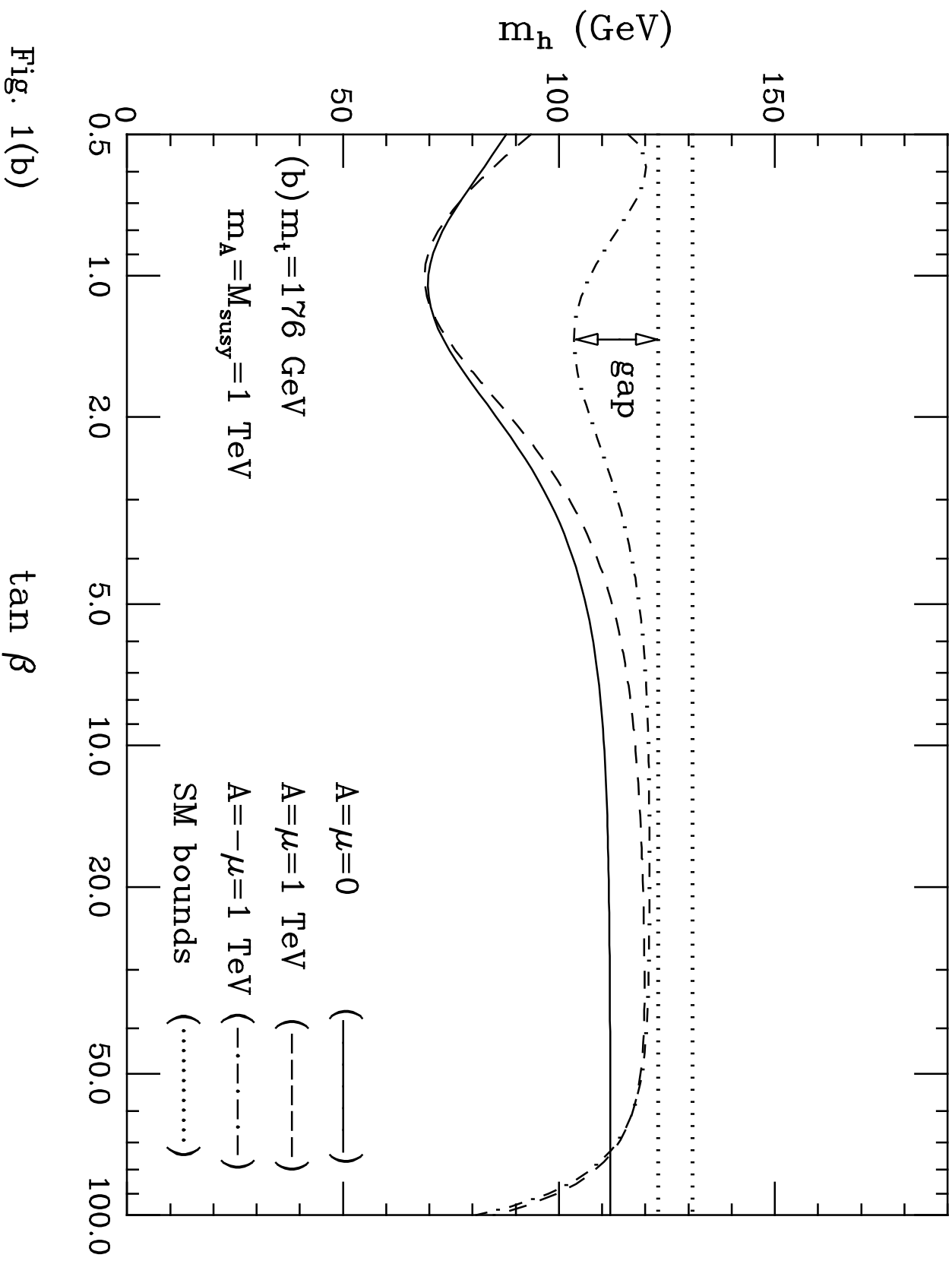
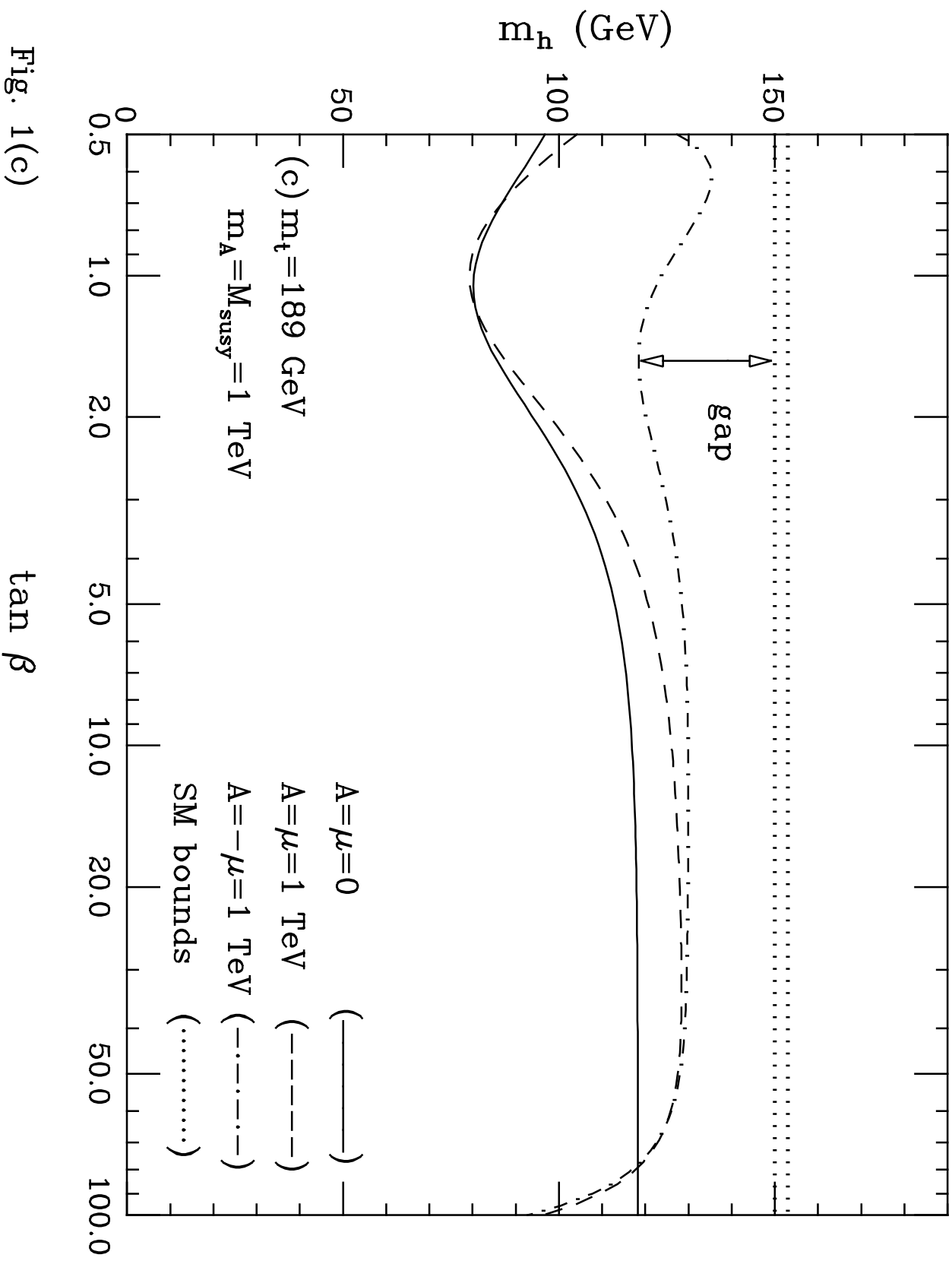
{\bf Fig. 1} The curves reveal the
upper bound on the lightest MSSM higgs particle
vs.\ $\tan\beta$, for top mass values of (a) 163 GeV,
(b) 176 GeV, and (c) 189 GeV.
Three extreme choices of susy parameters are invoked:
the solid curve is for $\mu=A_t=A_b=0$, the dashed curve is for
$\mu=A_t=A_b=1$ TeV, and the dot-dashed curve is for
$\mu=-1$ TeV, $A_t=A_b=1$ TeV.
In all cases, $m_A=m_{\tilde{q}}=1$ TeV and $m_b(M_Z)=4$ GeV are assumed.
The horizontal dotted lines are the ($\tan\beta$--independent) SM lower
bounds on the
higgs mass; the more restrictive stability bound derives from requiring
that the EW vev sits in an absolute minimum, while the less restrictive
metastability bound derives from requiring that the vev lifetime
in the local EW minimum exceed the age of the universe.
\\
\noindent
{\bf Fig. 2} \\
Upper bound on the lightest (M+1)SSM higgs vs.\ $\tan\beta$,
for the top mass values
(a) 163 GeV, (b) 176 GeV, and (c) 189 GeV.
All superparticles and higgses beyond the lightest are assumed to be heavy,
of order of the chosen susy--breaking scale of 1 TeV. The GUT scale is
taken as $10^{16}$ GeV.
\\
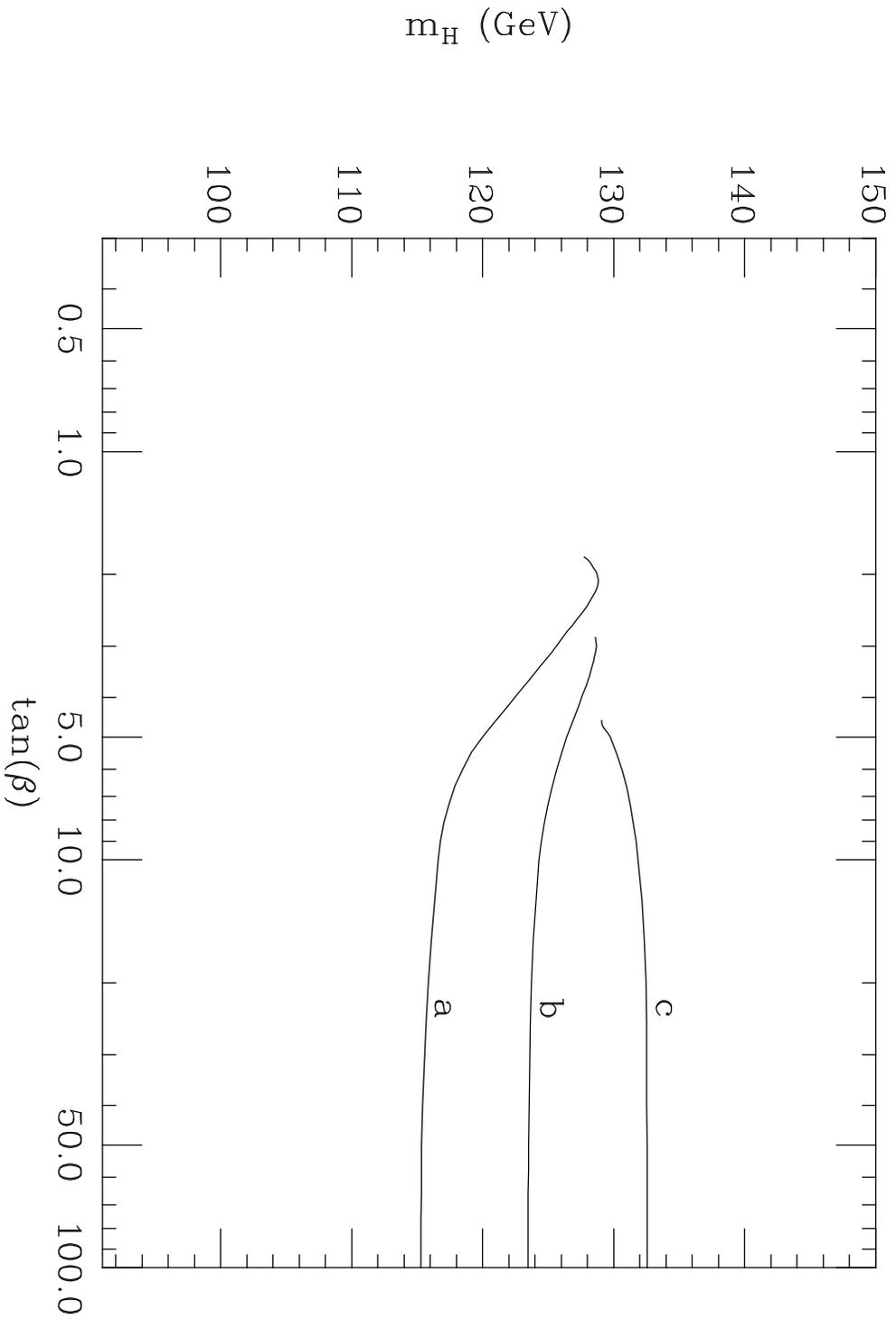
\end{document}